# TEDS: A Trusted Entropy and Dempster Shafer Mechanism for Routing in Wireless Mesh Networks

Heng Chuan Tan
School of Electrical and Electronic Engineering
Nanyang Technological University
Singapore
htan005@e.ntu.edu.sg

Houda Labiod
INFRES
Telecom ParisTech
Paris, France
Labiod@telecom-paristech.fr

Maode Ma
School of Electrical and Electronic Engineering
Nanyang Technological University
Singapore
emdma@ntu.edu.sg

Peter Han Joo Chong
School of Electrical and Electronic Engineering
Nanyang Technological University
Singapore
ehjchong@ntu.edu.sg

*Abstract*— **Wireless Mesh Networks (WMNs) have emerged as a key technology for the next generation of wireless networking due to its self-forming, self-organizing and self-healing properties. However, due to the multi-hop nature of communications in WMN, we cannot assume that all nodes in the network are cooperative. Nodes may drop all of the data packets they received to mount a Denial of Service (DoS) attack. In this paper, we proposed a lightweight trust detection mechanism called Trusted Entropy and Dempster Shafer (TEDS) to mitigate the effects of blackhole attacks. This novel idea combines entropy function and Dempster Shafer belief theory to derive a trust rating for a node. If the trust rating of a node is less than a threshold, it will be blacklisted and isolated from the network. In this way, the network can be assured of a secure end to end path free of malicious nodes for data forwarding. Our proposed idea has been extensively tested in simulation using network simulator NS-3 and simulation results show that we are able to improve the packet delivery ratio with slight increase in normalized routing overhead.**

## I. INTRODUCTION

Wireless Mesh Network (WMNs) are fast gaining popularity as the next generation of wireless networking due to their low setup cost, ease of implementation, good network coverage and self-management capabilities [1]. A WMN is made up of two types of nodes: the mesh routers and the mesh clients. The mesh routers are statically deployed and they form the wireless mesh backbone to provide network access for the mesh clients such as your laptops, smart phones or tablets, etc. The mesh clients on the other hand, can be static or mobile with simpler hardware and software requirements. Together, the mesh routers and the mesh clients cooperate to carry out packet forwarding via multi-hop communications to ensure proper data delivery.

However, due to the openness of the wireless medium and the multi-hop nature of communications in WMNs [2], we cannot assume all the nodes in the network are cooperative and well-behaved. Nodes may act selfishly by not forwarding the data packets in order to conserve their scarce resources, such as power and bandwidth. Second, the use of cryptographic techniques, although can deny unauthorized users access to the network, it may not be a viable solution as the mesh clients are resource limited. Also, if the nodes are compromised, they can retrieve the public and private keys used for communications and break the cryptography systems. Subsequently, they may conduct internal attacks by dropping packets to mount a Denial of Service (DoS) attack. If the compromised nodes drop 100% of the data packets, it is called a blackhole attack.

Several works have been proposed in literature [3]-[9] to deal with packet droppers or blackhole attacks. Zhang et al. [9] use a reputation driven mechanism called EigenTrust [10] to evaluate the trust of a node and integrates the trust information into an anomaly detection system to identify packet droppers in the network. This method, however, assumes the presence of prior trustworthy nodes which is not practical in WMNs because pre-trusted nodes may misbehave to protect their own interests. Proto et al. [7] use EigenTrust to compute the trust of a node via a path-wide approach. The trust values are transformed into a weighting metric in Optimized Link State Routing (OLSR) protocol [19] to determine the best path trustworthiness. One issue with this approach is that well-behaved nodes are treated unfairly. One misbehaved node in the forwarding path will result in the decrease of the reputation values of all other nodes along the path.

Marti et al. [6] proposed two mechanisms to detect misbehavior in Mobile Ad hoc NETworks (MANETs): Watchdog and pathrater. Watchdog uses overhearing technique to identify misbehaved neighboring nodes whereas pathrater is to keep state about the goodness of other nodes in order to decide the most reliable routes among the nodes. This method however, suffers from badmouthing attack as an attacker can malign a good node and cause other nodes to avoid the good node. Shila et al. [8] further extend the Watchdog mechanism by enabling both upstream and downstream traffic monitoring to enhance the detection capabilities in the presence of wireless losses. The disadvantage is that misbehaved nodes can only be detected by the source node based on the receipt of the PROBE ACK message. If PROBE ACK is not received by the source, then the source needs to initiate a hop by hop query for the PROBE and PROBE ACK packets which is going to increase the computational load.

In [3]-[5], the authors used subjective logic developed by Josang [11] to qualify trust where trust is represented by an opinion having belief, disbelief and uncertainty. The motivation behind this idea is that no one can determine with absolute certainty that a proposition is true or false. Hence, we can only form subjective opinion which contains certain degree of uncertainty regarding the truth of the proposition. It allows for better expressiveness and clarity than traditional probabilistic logic thereby allowing users to specify situations like "I don't know" or "I'm not sure". While subjective logic is effective in providing accurate opinions as it takes into account uncertainty, aggregation of trust opinions is complex and it requires high memory storage as each node needs to store the belief, disbelief and uncertainty parameters.

Motivated by the limitations of existing approaches, we propose a trust based mechanism called Trusted Entropy and Dempster Shafer (TEDS), through incorporating with Ad hoc On demand Distance Vector (AODV) routing protocol [19] to find a secure end to end path free of non-trusted nodes to safeguard against blackhole nodes.

The rest of the paper is organized as follows. Section II describes the threat model and assumptions. Section III presents the proposed TEDS design. Section IV presents the simulation results to demonstrate the performance of TEDS. Section V gives the conclusion and future works.

## II. THREAT MODEL AND ASSUMPTIONS

WMNs are exposed to security threats at any layer of the internet protocol stack which can cause the network to degrade or malfunction [13][14]. In this paper, we only focus on blackhole attacks at the network layer and state our assumptions for our TEDS design.

### A. Blackhole Attacks

In a blackhole attack, the malicious node will advertise itself as having the best route to the destination even though it does not have a route to it. It does this by sending a route reply (RREP) packet immediately to the source node [14]. The source node upon receiving this malicious RREP assumes the route discovery is complete and ignores all other RREP from other nodes and selects the path that now includes the malicious node as a relayer to forward the data packets. Subsequently, the malicious node will drop all the traffic received. The malicious node thus creates a blackhole in the network. An illustration is given in Fig. 1.

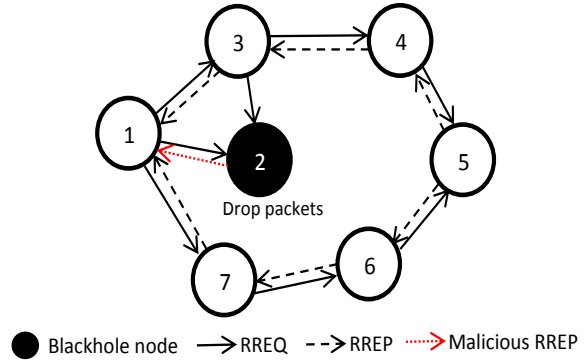

Figure 1. Blackhole Attack

### B. Assumptions

We assume the mesh routers and mesh clients in the network are statically deployed to model communications in infrastructure based WMNs and the majority of the mesh routers are cooperative and well-behaved. We assume that the sources and destinations are fully trusted. We further assume that the network is strongly connected where there are many other alternate paths from the source to the destination free of malicious nodes. Next, we assume the malicious nodes act independently and they do not collude. Therefore, our model is free of attacks, such as cooperative blackholes or wormhole attacks or any other forms of collusion attacks. The problem of colluding nodes is left as a future work. Lastly, trust value in our model is defined in the range $[0,1]$, where trust value of 0 means the node is untrusted and trust value of 1 means the nodes is fully trusted. Lastly, all nodes at the start of the network assume an initial trust value of 0.5.

## III. TEDS DESIGN

In this section, we present the details of TEDS. Lastly, we present our simulation settings with results and discussions.

### A. TEDS Overview

Our proposed model makes use of trust metric to assess the trustworthiness of a node in the network and it is based on three mechanisms. The three mechanisms are Watchdog [6], Shannon entropy function [12] and Dempster Shafer Theory [15]. In our scheme, we assume each node starts off with a trust value of 0.5 and each node is installed with Watchdog functionality to monitor the next hop forwarding behavior. The observations gathered from overhearing will be used to compute the forwarding probability of each neighbor node in the network. We then applied Shannon entropy function to compute the uncertainty of this forwarding probability thereafter, compute the direct trust value of each neighbor node. This direct trust value represents a node's direct experiences with its one hop neighbor. Next, we apply Dempster Shafer theory to combine conflicting trust values coming from a node's direct interactions as well as indirect interactions to determine the overall trust value. The indirect interactions in our context are defined as indirect trust derived from recommendation trust values from other nodes. This overall trust value will be calculated periodically *(every trust interval)* which will be feedback to the routing protocol for routing decisions.

### B. Calculation of Forwarding Probability

Each node is installed with Watchdog functionality to monitor the number of packets sent and the number of the packets overheard. To prevent the trustor *(node responsible for the evaluation of trust of its downstream node)* from falsely accusing its neighbor of misbehaving due to the inevitable collisions at the sender [6], we proposed that the trustor continues to monitor its downstream node for an extended period of time which we set to 2 seconds. If the trustor still fails to overhear the packet sent out by its neighbor after 2 seconds, the trustor concludes that the downstream node has dropped the packets maliciously. An example of how the watchdog works is given in Fig. 2. Each node keeps track of all sent packets by maintaining a table containing the ID of the node that the packet is directed to, the packet ID and the expiration time which is 2 seconds. When the trustor overhears a packet and finds a match in its corresponding table, the table entry for the overheard packet ID will be deleted. At every trust interval, we compute the forwarding probability, $f_p$ using (1).

$$f_p = \frac{\text{\# of overheard packets sent by n}_i}{\text{\# of packets send to n}_i \text{ for forwarding}} \quad (1)$$

### C. Calculation of Direct Trust

The next step is to formulate the direct trust value and we proposed using the Shannon binary entropy function defined in (2) together with a set of mapping equations defined in (3) to compute the direct trust values. This will ensure that the trust value is bounded by and confined in the range [0,1] and such that low forwarding probabilities correspond to lower trust values while high forwarding probabilities corresponds to high trust values.

$$H_b(p) = -p\log_2 p - (1-p)\log_2(1-p) \quad (2)$$

$$DT = \begin{cases} 1 - 0.5H(p), & \text{for } 0.5 \leq p \leq 1 \\ 0.5H(p), & \text{for } 0 \leq p < 0.5 \end{cases} \quad (3)$$

where $DT$ denotes the trust value of a node, $H_b(p)$ denotes the binary entropy function in (2) and $p$ denotes the forwarding probability derived in Section B.

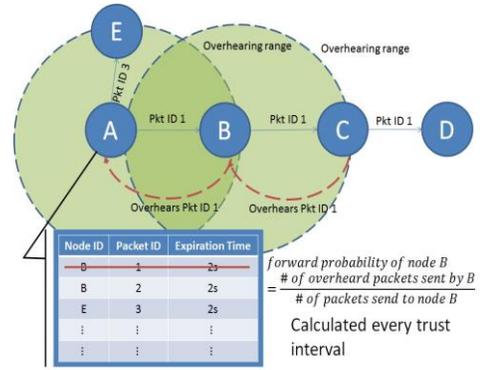

Figure 2. Watchdog Operation

### D. Update of Direct Trust

To better reflect the currency of direct trust values, we use Exponential Moving Weighted Average (EMWA) [16] to combine the past historical trust values of a node with the current measured trust values. EMWA applies exponentially decreasing weighting factors to each data point so as to smooth out the direct trust value and hence it provides a better representation of trust values over a period of time. EMWA for our trust model is given by (4).

$$DT_t = \alpha \cdot DT_t + (1-\alpha) \cdot DT_{t-1} \quad (4)$$

where $\alpha$ is a constant smoothing factor between 0 and 1, $DT_t$ represents the current trust value to be evaluated and $DT_{t-1}$ represents the previous trust value recorded by TEDS. If the smoothing factor $\alpha$ is large, it discounts the older observations faster. For our scheme, the smoothing factor $\alpha$ is selected to be 0.667. This means that a higher weightage is placed on a node's current trust value compared to the past. The choice of $\alpha$ is chosen to match the average age of data in simple moving average (SMA) and the formula is given by (5). The proof can be found in [17].

$$\alpha = \frac{2}{N+1} \quad (5)$$

where $N$ is the number of samples or records considered in EMWA.

### E. Calculation of Indirect Trust

Beside direct trust values, a node could also form an indirect trust value based on the recommendation trusts from other neighboring nodes, as shown in Fig. 3. Let us assume node A's trust of node B denoted by $DT_{AB}$ is 0.667 and node B's trust of C denoted by $DT_{BC}$ is 0.3. To determine the indirect trust value of node A about C taking into consideration node B's recommendation about node C, node B's trust of node C has to be weighted by node A's trust of node B according to (7). This is similar to transitive trust described in Eigentrust [10] and in subjective logic [11]. The purpose is to discount node B's direct trust of node C based on node A's observation of node B so as to mitigate the effects of badmouthing. Using (7), the weighted trust value *(indirect trust value denoted as IDT)* for the example given in Fig. 3 will be equal to 0.2.

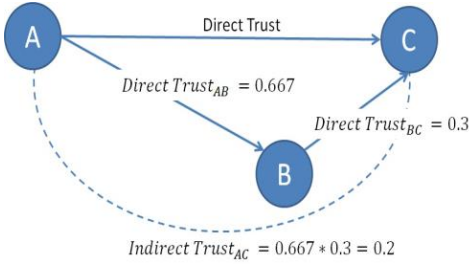

Figure 3. Formulation of Indirect Trust

$$IDT_{AC} = DT_{AB}.DT_{BC} \qquad (7)$$

### F. Combining Direct Trust and Indirect Trust

Dempster Shafer theory (DST) [15] is used to combine a trustor's direct trust of a node and indirect trust values received from other recommending nodes to arrive at the overall trust value of a node. DST is used because it allows us to quantify uncertainty in our trust computation instead of being forced to use prior probabilities to add up to 1 based on traditional probability logic. For instance, if the trust value of a node is 0.6, its complement probability which is distrust will be 0.4 according to traditional probability theory. Sometimes, it is unrealistic to make that claim because the lack of knowledge about an event is not regarded as evidence supporting the distrust of a node. Instead, DST classifies 0.4 as uncertainty which means a node can either be trusted or untrusted. Hence, DST can better reflect the behavior of the node and can improve on the trust evaluation.

First, nodes are classified into two states: Trusted ($T$) and Untrusted ($UT$). So the frame of discernment $\Theta$ in DST consists of $\{T, UT\}$. The power set denoted by $2^\Theta$ contains these four sets:

$$2^\Theta = \left[\{T\}, \{UT\}, \{T, UT\}, \{\phi\}\right] \qquad (6)$$

The set represented by $\{T, UT\}$ denotes uncertainty in DST, which means that a node can be trusted or untrusted. We apply the direct trust values we obtained from (4) and indirect trust values obtained from (7) as Basic Probability Assignment (BPA) to denote the strength of evidence pertaining to a particular subset of $2^\Theta$. In our scheme, trust value of 0.5 and above will be classified as trusted whereas trust value less than 0.5 will be classified as untrusted. For example, if the trust value of a node is 0.6, BPA for the set $\{T\}$ will be 0.6. The remaining belief mass of 0.4 will be allocated to the set $\{T, UT\}$. Following this rule and using Dempster's rule of combination in (8), direct trust and indirect trust can be combined to compute the overall trust value of a node.

$$m_{1,2}(S) = \frac{1}{1-K} \sum_{A \cap B = S \neq 0} m_1(A) m_2(B) \qquad (8)$$

where $K = \sum_{A \cap B = 0} m_1(A) m_2(B)$ is the normalization factor to ensure the total sum of combined masses, $m_{1,2} = 1$; $m_1(A)$ and $m_2(B)$ each represent the BPA assigned to direct trust and indirect trust, respectively.

### G. Decision Making

The overall trust value of a node is feedback to the routing protocol for routing decisions. If the overall trust value is ≥ 0.5, we conclude that node is trusted, else the node is misbehaving. If a node is detected as misbehaved, the trustor will add the misbehaved node to the blacklist. At the same time, it will broadcast a message throughout the network to inform other nodes of the blacklist nodes. Other nodes upon receiving the broadcast message will also put the misbehaved node in blacklist to avoid using it for all future communications. The trustor will also send a route error (RERR) message to notify the source node where another route discovery will be initiated to find a path free of malicious nodes.

## IV. SIMULATION

All our simulations are performed using Network Simulator NS3 (v3.17) [18]. TEDS is integrated into AODV [20] of NS-3 [18]. Although AODV is used, our proposed trust system is independent of the underlying routing protocols. It is compatible with other routing protocols as it runs on top of any routing protocol. TEDS only triggers the underlying routing protocol to re-initiate a new route discovery upon detection of malicious nodes. To evaluate the performance of TEDS, we compared it with basic AODV that is without any trust mechanisms.

### A. Simulation Environment

Our simulation environment consists of 100 nodes placed in a square grid manner. The distance between each adjacent node is 150m and the radio range of each node is 250m. The source and destination nodes are located on the left and right side of the square grid. All nodes in our simulation environment are assumed static to model the WMN backbone infrastructure. We simulate a total of 10 CBR flows between randomly selected source nodes on the left and randomly selected destination nodes on the right.

The maximum packet per flow is configured as 300 with a packet generation rate of 4 packets/s. The start time of each flow is uniformly distributed between 30 seconds and 200 seconds. A random number generator is used to randomly select the source and destination nodes as well as to locate the malicious nodes in the forwarding path between each source and destination pair. Simulations were performed for a period of 300s and each data point is an average of 10 runs unless otherwise stated. More simulation parameters are given in Table I.

TABLE I. SIMULATION PARAMETERS

| Simulation tool | NS-3 |
|---|---|
| Grid spacing | 150m |
| Data rate | 16kbps |
| Transmission range | 250 m |
| Network area | 1500 m x 1500 m |
| No: of nodes in the network | 100 |
| Traffic | 10 source-destination pairs |
| Packet size | 512 B |
| Packet generation rate | 4 packets/s |
| Simulation time | 300 s |
| Traffic type | CBR |
| Transport protocol | UDP |
| Routing protocol | AODV (disable HELLO) |
| Mac protocol | IEEE 802.11b |
| Propagation loss model | RangePropagationLossModel |
| Physical layer | YansWifiPhy channel |
| Mobility | Static |

The performance of TEDS is evaluated for the following cases. First, we study the performance when the network is under blackhole attack. Next, we examine the sensitivity of the trust interval on the performance of TEDS in terms of packet delivery ratio and routing overhead. The selection of trust interval determines how often TEDS computes and propagates the trust value in the network. For all these experiments, we made two assumptions. We assume that all source and destinations are trustworthy and the second assumption is that all nodes start off with a default trust rating of 0.5 at the initial state.

*B. Results and Discussions*

*1) Performance Analysis under Blackhole Attacks*

We first evaluate the performance of TEDS under the effects of blackhole attacks and we compare the result with basic AODV without any trust mechanisms. We are interested in the Packet Delivery Ratio (PDR) and normalized routing overhead of TEDS compared to basic AODV. To simulate the blackhole attacks in NS-3, we configure the malicious nodes to send a RREP packet with a high sequence number such that it will be selected by the source node during route discovery. Subsequently, the malicious nodes will drop 100% of the data packets. The comparison was done by varying the number of blackhole nodes in our network and the blackhole nodes are randomly selected from the network area of 80 nodes discounting the source and destinations which are on opposite sides of the square grid.

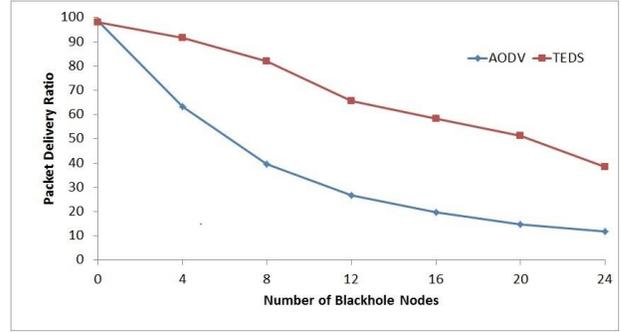

Figure 4. PDR performance in the presence of blackhole nodes

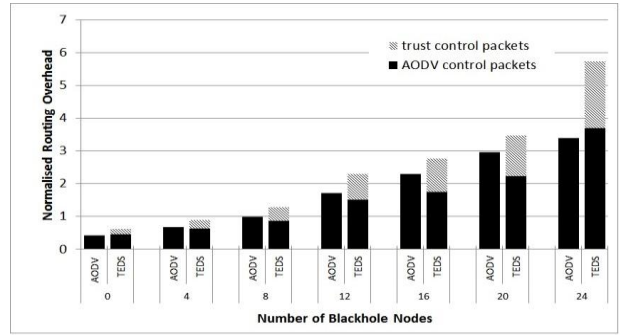

Figure 5. Normalized routing overhead in the presence of blackhole nodes

Fig. 4 shows the average PDR of TEDS compared with basic AODV. As shown in Fig. 4, the PDR decreases when the number of malicious nodes increases. When there are no malicious nodes in the network, we are able to achieve similar PDR performance for TEDS and AODV. As the number of malicious nodes increases, the PDR of TEDS decreases at a slower rate but still able to achieve about 30% improvement on the PDR compared to AODV. This shows that TEDS is capable of detecting and isolating malicious nodes using the trust mechanisms introduced. As more and more blackhole nodes are introduced into the network, TEDS's performance also starts to decrease gradually; this is because as more blackhole nodes are being identified and isolated by TEDS, there remain fewer alternatives available for choosing the forwarding paths considering our network is a static environment.

Fig. 5 compares the normalized routing overhead for TEDS and basic AODV when the network is under blackhole attacks. Based on Fig. 5, two observations can be made. First, the normalized routing overhead incurred by TEDS is higher than AODV. Second, the normalized routing overhead increases with increasing blackhole nodes in the network. The increase in normalized routing overhead is due to the following reasons: (1) periodic exchange of trust information with neighboring nodes, (2) broadcast of trust control message to notify other nodes of blackhole nodes, so that they can isolate them and not use them for packet forwarding and (3) re-initiation of a new route discovery upon detection of blackhole nodes. We further show the breakdown of the normalized routing overhead for TEDS into control packets introduced by our trust mechanism and control packets introduced by AODV. We conclude that the increase in normalized routing overhead for TEDS is mainly due to the trust related control packets which also increases proportionally with the increase of blackhole nodes in the network. This increase in overhead is in the acceptable range in exchange for higher security and higher PDR.

*2) Selection of Trust Interval*

In this experiment, we assess the sensitivity of the trust interval on the performance of TEDS. The trust interval determines how frequent we perform the trust computations and propagation of trust values in the network. We compare the PDR vs. trust interval with 10% blackhole nodes in the network. The trust interval is varied from 10s to 60s and each data point on the plot represents an average of 20 runs.

Fig. 6 shows the PDR for TEDS by varying the trust interval period. The curve shows that the PDR is the highest when trust interval is 10s. Beyond 10s, the PDR starts to drop. The PDR is around 85% at trust interval of 10s compared to 60% when the trust interval is configured as 60s. One reason is that, when trust query period is small, a malicious node will be detected earlier and hence less number of packets will be lost. As trust interval increases, more packets are lost due to the dropping behavior of the malicious nodes and that attributed to the drop in PDR as trust interval increases.

Next, we examine the effects of trust interval on the routing overhead. The routing overhead in our case consists of control packets generated by TEDS (trust-related control packets) and AODV related control packets for route discovery and route maintenance (RREQ, RREP, RERR). From Fig. 7, we see that the number of routing control packets is the highest when the trust interval is at 10s and that it tapers off as the trust interval increases. The reason is that at smaller trust interval, nodes need to exchange and disseminate trust related control packets more frequently which resulted in the increase in routing overhead. Based on Fig. 6 and Fig. 7, we can have the following observation. There is a trade-off between PDR and routing overhead vs. the trust interval. High PDR is achieved when the trust interval is small but at smaller trust interval, the routing overhead is high which makes TEDS less efficient and more costly in resources. Here, we conclude that the optimum trust interval based on simulation is 20s.

## V. CONCLUSION AND FUTURE WORK

In this paper, we proposed a novel lightweight trust mechanism called TEDS to detect blackhole nodes in the forwarding path of a node. TEDS exploits the broadcast nature of wireless medium to listen promiscuously to the next node's transmission to detect packet droppers. It uses Shannon entropy to derive the direct trust value of a node and we demonstrated how to combine direct trust and indirect trust to form a shared belief using Dempster's rule of combination. Benefits of TEDS are that it is lightweight; uncertainty is quantified in the trust computations and it is portable. TEDS can be integrated into any routing protocols. Our simulation results show that TEDS can detect packet droppers and improved the packet delivery ratio of the network as the number of malicious nodes increases with minimum increase in normalized routing overhead. We further show that the optimum trust interval is 20s through simulation where TEDS is able to ensure high PDR with reasonable routing overhead. For future work, we plan to study collusion attacks such as cooperative blackhole or wormhole initiated by multiple attackers and to study the impact of node mobility on the performance of TEDS.

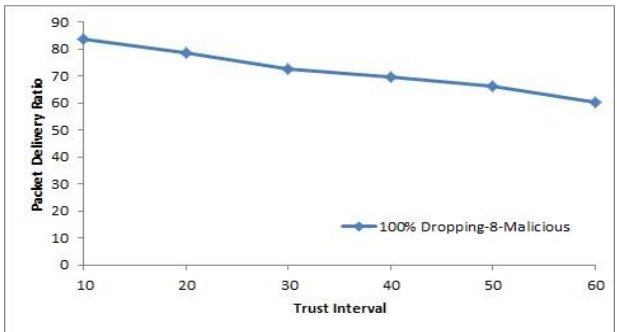

Figure 6. PDR performance under varying trust interval

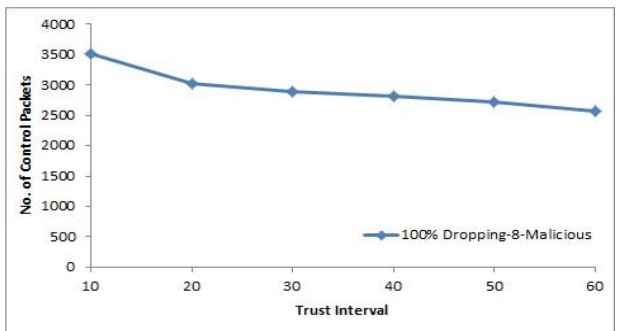

Figure 7. Routing overhead under varying trust interval